%% file: arxiv08anonymization.tex
\documentclass[10pt]{article}
\pdfoutput=1
\usepackage{fullpage}
\usepackage{url}
\usepackage{graphicx}

\newcommand{\alertsa}{\ensuremath{\mathcal{A}_{anony}}}
\newcommand{\alertsb}{\ensuremath{\mathcal{A}_{baseline}}}
\newenvironment{titemize}{
\begin{itemize}
  \setlength{\itemsep}{1pt}
  \setlength{\parskip}{0pt}
  \setlength{\parsep}{0pt}}{\end{itemize}
}
\newenvironment{tdescription}{
\begin{description}
  \setlength{\itemsep}{1pt}
  \setlength{\parskip}{0pt}
  \setlength{\parsep}{0pt}}{\end{description}
}
\title{Evaluating the Utility of Anonymized Network Traces for
  Intrusion Detection}
\author{\begin{minipage}{3in}
\centering
Kiran Lakkaraju \\
Department of Computer Science \\
University of Illinois, Urbana-Champaign \\
\end{minipage}
\begin{minipage}{4in}
\centering
Adam Slagell \\
National Center for Supercomputing Applications \\
University of Illinois, Urbana-Champaign
\end{minipage}}
\begin{document}
\maketitle

\input{abstract}
\input{intro}

\input{flaim}

\input{method}

\input{results2}

\input{multifieldresults}

\input{related}

\input{conclusion}

\input{appendix}

\end{document}

%% file: abstract.tex
\begin{abstract}

Recently, it has become increasingly important for computer security
researchers and incident investigators to have access to larger and
more diverse data sets. At the same time, trends towards protecting
customer privacy have grown as a result of many embarrassing
releases---of supposedly anonymous information---which has been traced
back to individual computer users \cite{aol06nytimes,
  narayanan06howtobreak}. This has further increased reluctance of
data owners to release large data sets to the research community or to
share logs relevant to attacks from a common threat. The burgeoning
field of data sanitization has helped alleviate some of these problems
as it has recently provided many new tools for anonymizing sensitive
data, but there is still a difficult trade-off to be negotiated
between the data owner's need for privacy and security and the
analyst's need for high utility data. Data sanitization policies must
be created that are secure enough for the first party, but do not
result in too much information loss to be usable to the second.

Necessary to solving this problem of negotiating policies for data
sanitization is the ability to analyze the effects of anonymization on
both the security of the sanitized data and the utility left after
anonymization. In this paper, we focus on analyzing the utility of
network traces post-anonymization. Of course, any such measure of
utility will naturally be subjective to the type of analysis being
performed. So this work scopes the problem to utility for the task of
attack detection. We employ a methodology we developed that analyzes
the effect of anonymization on Intrusion Detection Systems (IDS), and
we provide the first rigorous analysis of single field anonymization
on IDS effectiveness. This work will begin to answer the questions of
whether the field affects anonymization more than the algorithm; which
fields have a larger impact on utility; and which anonymization
algorithms have a larger impact on utility.

\end{abstract}

%% file: intro.tex
\section{Introduction}\label{intro}
The ability to safely share log files and network traces has become
increasing important to several communities: networking research,
computer security research, incident response, and education
\cite{slagell05sharingcomputer}. Synthetically generated data is
abundant, but has been highly criticized for many uses
\cite{mchugh00testing}. Honeynets can be useful in generating
exercises for students and help meet the needs of educators
\cite{honeynetChallenge}, but very few honeynets have been setup on a
scale to generate some of the large, cross-sectional data sets needed
by computer security researchers
\cite{vrable05scalability}. Furthermore, using a honeynet does not
necessarily release the owner from all legal responsibility when
sharing data \cite{sicker07legalissues}. Lastly, nothing but real data
will suit the needs of the incident responder who must share data
about specific attacks under investigation of real
machines. Therefore, there is a high demand for methods to share {\em
  real} log files and network traces within several communities.

At the same time as there is an increased need to share these data
sets, there is increased reluctance. First, many data owners recognize
the inherent security risks of releasing detailed information that
could be used to map out their own networks, services or sensor
locations
\cite{bethencourt05mappinginternet,coull07playing,kohno05remotephysical,lincoln04privacypreserving,pang06devil,Pang03,xu02prefixpreserving,zalewskip0f}. Secondly,
there are serious privacy concerns that companies have about releasing
customer data, especially in light of recent incidents where some
publicly released data---believed to be anonymous---leaked information about
specific users \cite{aol06nytimes,narayanan06howtobreak}. Lastly,
recent research has questioned the legality of releasing much of the
data as has been done up until now \cite{sicker07legalissues}.

The burgeoning field of data sanitization has helped address this
tension by providing organizations, who wish to share their data, with
new tools to anonymize computer and network logs
\cite{coull07playing,foukarakis07flexible,lundin99privacy,pang06devil,seeberg07anewclassification,slagell06flaim,slagell05sharingnetworklogs,slagell05networkloganonymization,slagell05networkloganonymization,xu02prefixpreserving,yurcik07anonyids}. However,
little has been done to help users negotiate the difficult trade-off
between the data owner's need for security and privacy, and the data
analyst's need for high quality data---what is called the {\em
  utility~vs.~security~trade-off} \cite{slagell06flaim}. As
anonymization is an inherently lossy process, and the data analyst
wants information as close to the original as possible, there is
always this tension and a need to negotiate policies that meet the
needs of both parties.

Necessary to solving this problem of negotiating policies for data
sanitization is the ability to analyze the effects of anonymization on
both the security of the sanitized data and the utility left after
anonymization. In this paper, we focus upon the latter problem of
evaluating the effects of anonymization on the utility of the data
sets to be shared. Of course, utility is subjective since it depends
upon who is using the data, or more specifically, for what purpose it
is being used. Hence, what is important to a researcher in the network
measurements community may be completely irrelevant to the incident
responder. Therefore, we have scoped this work to evaluating utility
for attack detection.

The task of attack detection is an important part of the incident
responder's daily job. When investigating broad attacks, of which
their organization is only a part, they may have to settle for
anonymized logs from the other sites involved. It would be a similar
case when using a distributed or collaborative intrusion detection
system that crosses organizational boundaries. Output from the sensors
may need to be anonymized. Not only is attack detection important in
these ``real world'' applications, but it is important to the
intrusion detection research community, as well. This community has
often complained that their only good data sets to test new
technologies against are synthetic. However, if they can still do their
analysis with anonymized data, then they are more likely to obtain
large, real data sets.

The utility of a data set is not only constrained by the type of
analysis being done with it but also the type of data being shared. We
have chosen to look at the effects of anonymization on one of the most
commonly shared and general types of data, the pcap formatted network
trace. From this type of data, many others can be derived (e.g.,
NetFlows) \cite{argus08website}.

To quantitatively measure the ability to identify attacks in
anonymized data, we developed what we call the {\em IDS Utility
  Metric}. This measurement evaluates and compares the false positive
and negative rates of a baseline {\em unanonymized} data set with that
of an {\em anonymized} data set. By doing this, we can automate an
objective process to help us answer several questions: (1) How does
anonymization of a particular field affect the ability to detect an
attack, (2) are there unique effects when certain pairs or triplets of
fields are anonymized together, and (3) how does the use of different
types of anonymization algorithms affect attack identification? In
this paper, we present the results of anonymizing a portion of the
1999 MIT/Lincoln Labs DARPA data set---by using the FLAIM
\cite{slagell06flaim} anonymization framework---with over 150 separate
anonymization policies to help us answer these questions.



The rest of the paper is organized as
follows. Section~\ref{sec:flaim:-framework-log} describes the
anonymization algorithms used by FLAIM in our experiments, while
section~\ref{sec:methodology} describes the methodology and
setup of our experiments. In section~\ref{sec:results-analysis}, we
present our results and analysis. We survey the related work in
section~\ref{related}, and state our conclusions, as well as scope out
future work to be done, in sections~\ref{sec:conclusions}
and~\ref{sec:future-work}.

%% file: flaim.tex
\section{FLAIM: Framework for Log Anonymization and Information Management}
\label{sec:flaim:-framework-log}

We chose to use FLAIM \cite{slagell06flaim} as our anonymization
engine for several reasons. First, we could easily script its
execution for a multitude of tests. Second, it has a very flexible XML
policy language that made it simple to generate hundreds of unique
anonymization policies. Third, it can anonymize as many or more fields
in PCAP traces as any other anonymization tool. Lastly, FLAIM has a
very rich set of anonymization algorithms that can be applied to all
these fields. With these properties, it was the ideal tool for our
experiments.

\subsection{Anonymization Algorithms} 
FLAIM implements a plethora of anonymization algorithms for several
data types. The three basic data types available to most anonymization
algorithms are \emph{binary, string,} and
\emph{numeric}. Additionally, there are a few special data types like
{\em timestamps}. Binary data is just treated as a string of bits
with no special structure. Algorithms anonymizing binary data
output binary data of the same length. String data are variable
length, terminated by a null character. Anonymization algorithms
that take in strings will also output stings. However, the
length may change. Numeric data is interpreted as a number of a
given base, specified in the anonymization policy. This is
useful, for example, when working with decimal numbers like a
port number. There, one may want to act on individual digits,
rather than bits (e.g., replacing the last 3 digits with 0's).

Table~\ref{tab:anony-algorithms} lists the different anonymization
algorithms in FLAIM along with the data types they operate
upon. Further information on these algorithms can be found in
\cite{flaim-manual}.

\subsection{Anonymization Policies in FLAIM}
FLAIM provides an expressive and powerful method for specifying
anonymization policies that can be modified at run time, thus
enabling efficient automation. An anonymization policy is an XML file
that specifies the anonymization algorithms that should be applied to
the various fields in the log, along with any special parameters to be
passed to those algorithms.

%% file: method.tex
\section{Methodology}\label{sec:methodology}

\begin{figure}[tp]
  \includegraphics[width=2in]{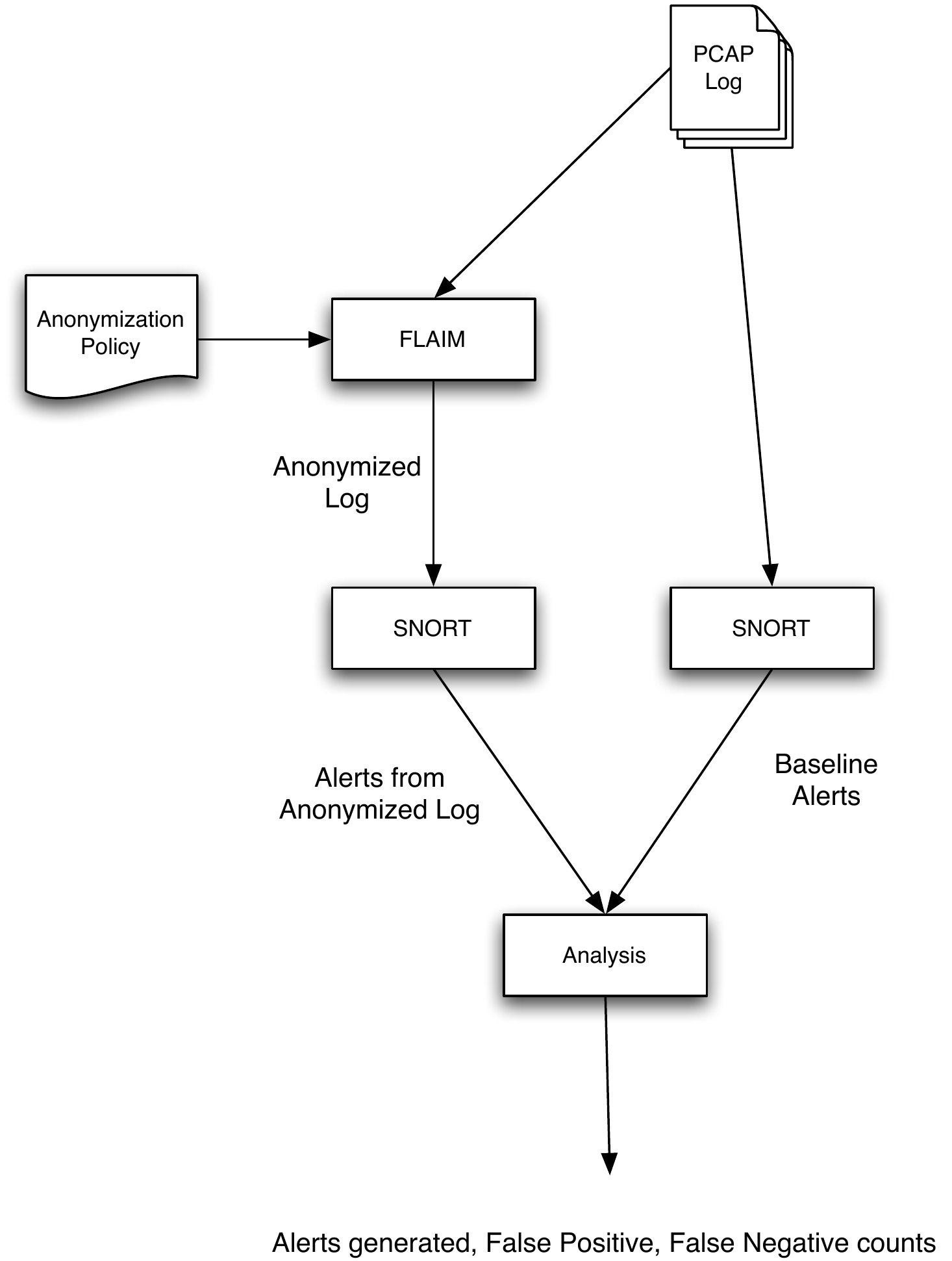}
  \caption{This diagram illustrates the process by which we compare anonymized logs versus non-anonymized logs.}
  \label{fig:methodology}
\end{figure}

For all of our experiments, we used a subset of the 1999 DARPA
evaluation data set. The Defense Advanced Research Projects Office
(DARPA) created an Intrusion Detection Evaluation testbed in 1998 and
1999. Data was captured from a simulated network that was subjected to
various attacks. This data set has been frequently used in evaluating
intrusion detection systems since its creation
\cite{lippmann00darpa1999}. Thus, we found it appropriate to use in
evaluating the effects anonymization of data can have on intrusion
detection.

As we mention, we used but a portion of the 1999 data set. Specifically, we
used the {\em inside} tcpdump data from Wednesday of the second week of the
evaluation. Since FLAIM currently just anonymizes TCP, UDP and ICMP,
we filtered out all other network protocols before running our
experiments, whose methodology is depicted in Figure~\ref{fig:methodology}.


The {\em IDS Utility Metric} is constructed as follows. First, the
unanonymized data set is processed by Snort to produce the
``baseline'' set of alerts, using the default Snort rule sets. We
consider this the ideal set of alerts for the data set. Next, we take
the same data set, and anonymize it---in our case, with FLAIM. We then
take the anonymized data set, and we run Snort against it with the
same rule set as before. The alerts generated from the unanonymized
file are used as a baseline against which the alerts generated by the
anonymized file are compared. The difference between the alerts in the
anonymized file, versus the unanonymized file, is used as a measure of
the loss of utility in the log. The larger the difference, the more
information was not available in the logs in order for the IDS to
correctly identify an attack. This process is depicted in
Figure~\ref{fig:methodology}.

FLAIM schemas specify a set of anonymization algorithms that are
appropriate for each field in a pcap log. These are summarized in
Figure~\ref{fig:pcapanoy}. In this evaluation, we only consider
anonymization policies that transform single fields. Before we can
evaluate the affect of multi-field policies, we must come to an
understanding of single field policies. There are 152 single field
policies that can be generated for pcap data
(Figure~\ref{fig:pcapanoy} in the Appendix lists all the fields and
anonymization algorithms used for testing). Each anonymization
algorithm also has parameters that affect how it anonymizes the
field. We will not go over the parameters in detail, but instead refer
readers to the FLAIM manual
\cite{flaim-manual}. Table~\ref{tab:anonyparam} in the Appendix 
summarizes the parameter settings for the anonymization algorithms.

We iterate the process described above over each of the 152
anonymization policies, comparing the results pre- and
post-anonymization. We describe how we compare these data sets in the
next section, while the section after discuses the actual metric in
more detail.

\subsection{Comparing Snort Alerts}
Alerts generated by Snort are defined by several properties (a full
list is included in the appendix as
table~\ref{tab:snort-alert-fields}). Each alert is associated with a
specific packet. The relevant alert fields---for our purposes---are
listed below:

\begin{description}
  \item[timestamp]: the timestamp from the offending packet.
  \item [sig\_generator]: the part of Snort generating alert.
  \item[sig\_id]: the Id. number of the signature that was fired.
  \item[msg]:  description of the alert.
  \item[proto]: the protocol of the offending packet.
  \item[src]: the source IP address of the offending packet.
  \item[srcport]: The source port of the offending packet.
  \item[dst]: the destination IP address of the offending packet.
  \item[dstport]: the destination port of the offending packet.
  \item[id]: Packet Id.
\end{description}

To determine whether two alert sets are equal we need a way of
determining if two alerts are equal. Normally this can be done by
comparing each field of the alerts. However, in this case the alerts
generated from the anonymized log will cause alerts that are actually
equal to appear unequal. To overcome this, we compare alerts on fields
which will not change due to anonymization. This leads to two distinct
field sets that must be used when comparing alerts. They are shown in
Table~\ref{tab:alert-field-sets}. Field Set 1 is used when the
timestamp field has not been anonymized. Field Set 2 is used when the
timestamp field has been anonymized.

\begin{table}
  \centering
  \begin{tabular}[t]{c|c}
    Field Set 1 & Field Set 2 \\ \hline
    timestamp & sig\_id \\
    sig\_id & src \\
    id & srcport \\
    & dst \\
    & dstport \\
    & id \\
    & tcpseq \\
  \end{tabular}
  \caption{The two field sets that are used to compare alerts.}
  \label{tab:alert-field-sets}
\end{table}

\subsection{Metrics for evaluating utility}

The purpose of anonymization is to share logs while hiding sensitive
information. Anonymization, while inherently an information reducing
procedure, must be measured in terms of the amount of information that
is lost in the file.  However, by anonymizing we can introduce new
false patterns into the data. The ``best'' anonymization policy should
minimize information loss, while not adding any new false patterns to
the log.

We can consider the IDS process as a pattern classification
process. The data set is input, and the IDS classifies each packet as
malicious or not. The alerts generated from the unanonymized data
(which we call the baseline data) are considered to be the correct
analysis of the data set. We then compare the alerts generated by the
anonymized data to the baseline data alert set.

Let the set of alerts generated from the baseline file be called
\alertsb ~and the set of alerts generated from the anonymized file be
called \alertsa. In terms of alerts we should compare \alertsa
vs. \alertsb. The best result would be for the anonymized alerts to
match, exactly, the alerts generated in the baseline set. Let us
consider the baseline alerts to be the target set. The alerts from the
anonymized file will be the generated set. Then we can define several
metrics:

\begin{description}
\item[True Positive] $TP = \alertsa \cap \alertsb|$ The number of alerts in \alertsa that are also in \alertsb.
\item[False Positive] $FP = |\alertsa - (\alertsa \cap \alertsb)|$ The number of alerts that were generated by the anonymized file, but were not in the baseline file.
\item[False Negative] $FN = |\alertsb - (\alertsa \cap \alertsb)|$ The number of alerts that were not caught by the anonymized file.
\end{description}

The True Positive rate indicates how much of the information was
preserved in the anonymized log. The False Positive rate indicates how
many additional patterns were added to the log through
anonymization. The False Negative rate indicate the amount of
information that was removed from the log. A good anonymization policy
should make sure both False Positive and False Negative are low while
maximizing the True Positive rate.

While the false positive rate is an important factor of primary importance is the false negative rate, as it indicates the loss of information through anonymization. For the remainder of this paper, we use both False Positive and False Negative rates as a measure of the utility of a log post-anonymization, but focus more on the false negative rate.

%% file: results2.tex
\section{Results and Analysis}
\label{sec:results-analysis}

In this section, we describe the results of our experiments with single
field anonymization policies. These experiments provide a substantive
start to answering these questions:

\begin{itemize}
\item What affects utility more, the fields that are anonymized or the anonymization algorithm?
\item Which fields have a larger impact on the utility of a log?
\item Which anonymization algorithms have a larger impact on the utility of a log?
\end{itemize}

To answer these questions, we evaluated all 152 pairs of fields and
anonymization algorithms. For each pair, the number of alerts
generated by Snort was calculated. The alerts generated for each pair
were compared with the baseline alerts (See
Table~\ref{table:raw-anony1} and Table~\ref{table:raw-anony2}). False
Positives/False Negatives were calculated based on the definitions
above, whose detailed results we discuss later.


The unanonymized file produced 81 alerts. The number and types of
alerts produced are summarized in
Table~\ref{table:baseline-alerts}. There are a total of 81 alerts
generated in the baseline file, but only 19 unique types of alerts.

\input{rawAlertData.tex}

\begin{table}[htp]
 \caption{Alerts generated in the baseline unanonymized file. \textbf{AlertID} is the id of the alert; \textbf{Num} is the number of that type of alert generated; \textbf{Desc} is a description of the alert}
  \label{table:baseline-alerts}
  \begin{tabular}[t]{llp{6in}}
    {\textbf{AlertID}} & {\textbf{Num}} & \textbf{Desc} \\
    1 & 1 & (portscan) TCP Portscan \footnote{This alert was generated from generator 122} \\
    323 & 1 & FINGER root query \\
    330 & 1 & FINGER redirection attempt \\
    332 & 1 & FINGER 0 query \\
    356 & 1 &  FTP passwd retrieval attempt \\
    359 & 1 & FTP satan scan \\
    503 & 4 & MISC Source Port 20 to <1024 \\
    1200 & 9 & ATTACK-RESPONSES Invalid URL \\
    1201 & 12 & ATTACK-RESPONSES 403 Forbidden \\
    1288 & 4 &  WEB-FRONTPAGE /\_vti\_bin/ access \\
    1292 & 30 &  ATTACK-RESPONSES directory listing \\
    1418 & 2 & SNMP request tcp \\
    1420 & 2 &  SNMP trap tcp \\
    1421 & 1& SNMP AgentX/tcp request\\
    2467 & 1 & NETBIOS SMB D\$ unicode share access \\
    2470 & 1 & NETBIOS SMB C\$ unicode share access \\
    2473 & 1 & NETBIOS SMB ADMIN\$ unicode share access \\
    3151 & 6 &  FINGER / execution attempt \\
    3441 & 2 & FTP PORT bounce attempt \\
  \end{tabular}
\end{table}

\subsection{Fields or Anonymization Algorithms?}
It is important to understand which causes a greater impact on utility; the field that is being anonymized or the anonymization algorithm that is being applied.

To evaluate the effects of anonymizing a field we calculate the
\emph{marginal} of a field. The marginal of a field is the average
number of false positive/false negatives over all anonymization
algorithms. Similarly, the marginal of an anonymization algorithm is
the false positive/false negative rate averaged over all fields. The
marginal provides a concise summary of the effect of anonymizing a
particular field or using a particular anonymization algorithm.

\begin{figure*}[htp]
  \begin{tabular}{cc}
    \includegraphics[width=3in]{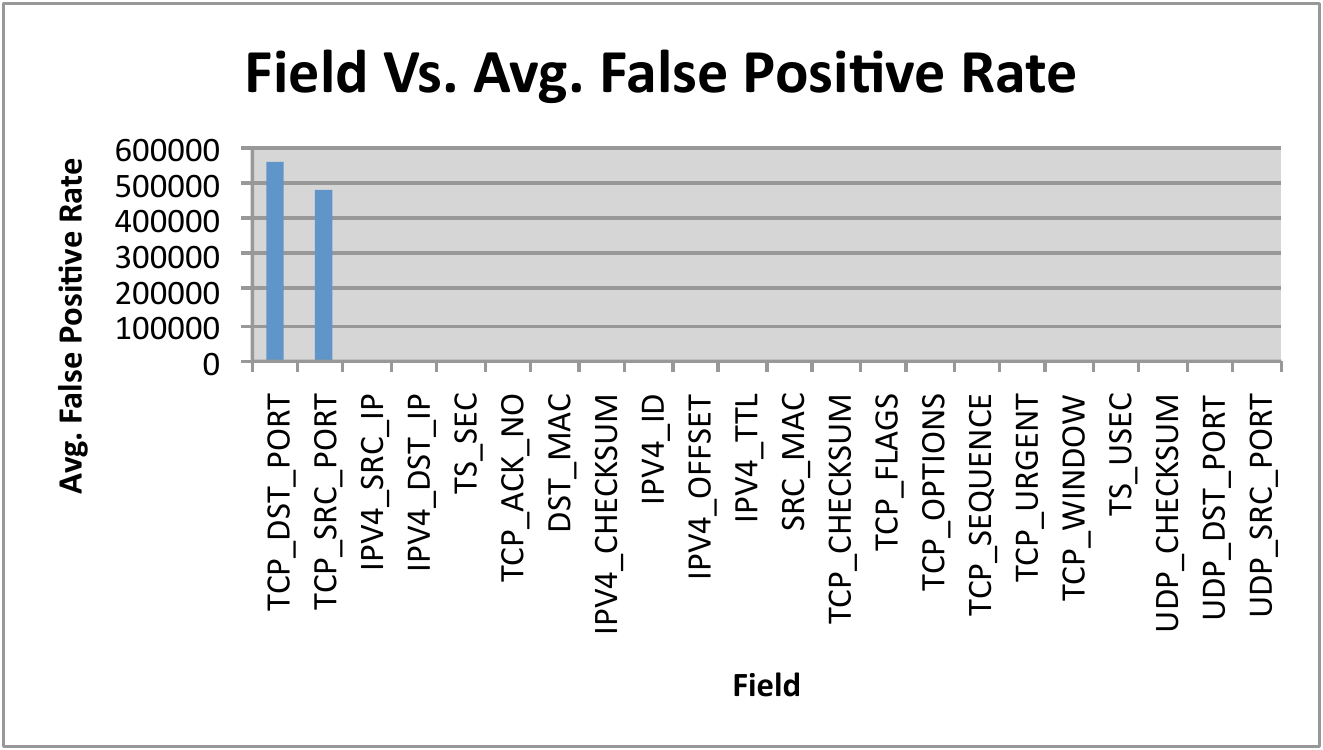} & \includegraphics[width=3in]{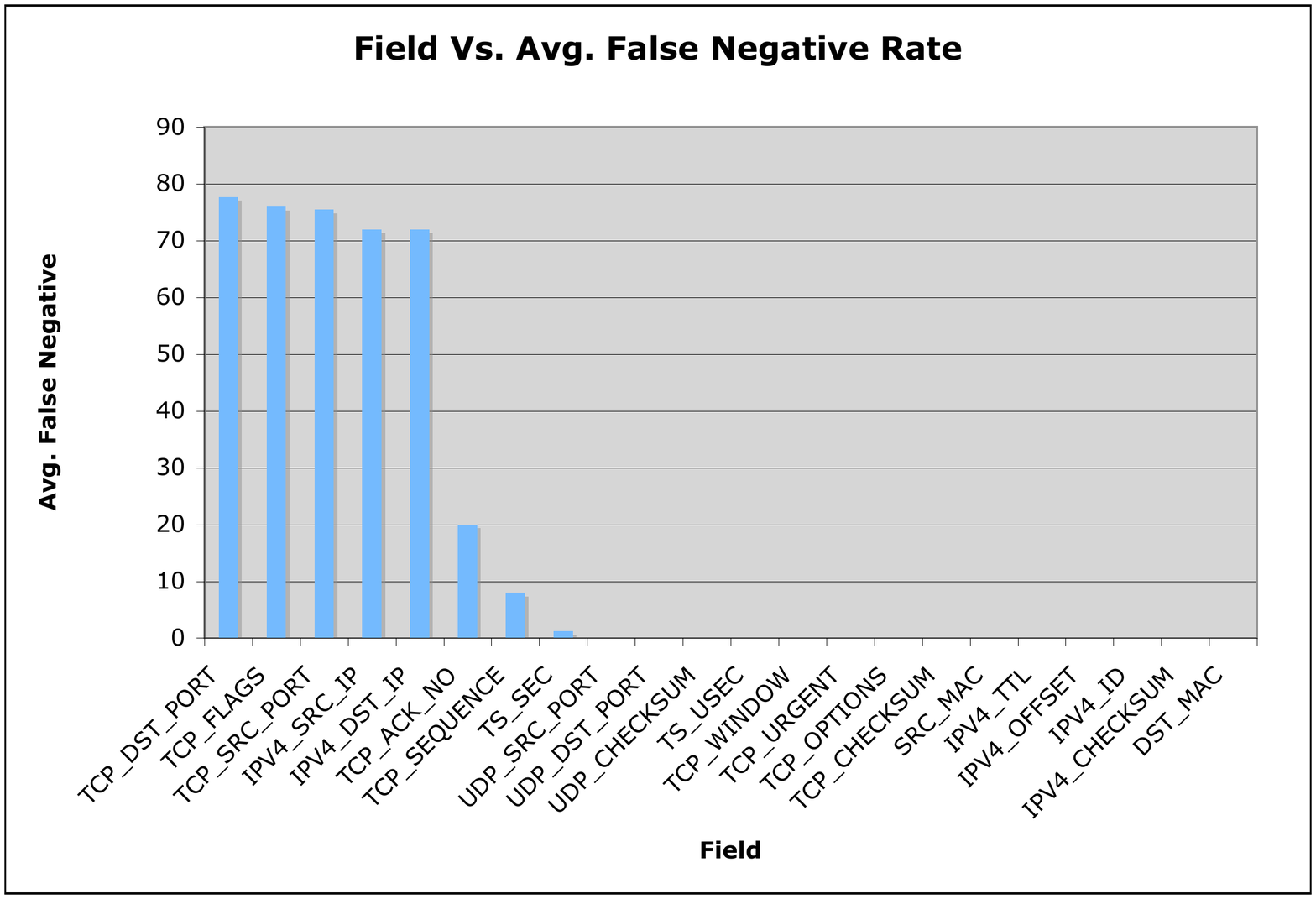} \\
  \end{tabular}
  \caption{The left hand chart shows the marginal of each field with respect to false positives (i.e, the average number of false positives for a field, averaged over all anonymization algorithms). The right hand chart shows the marginal of each field with respect to false negatives.}
  \label{fig:marg-fieldvsfpfn}
\end{figure*}

\begin{figure*}[htp]
  \begin{tabular}{cc}
  \includegraphics[width=3in]{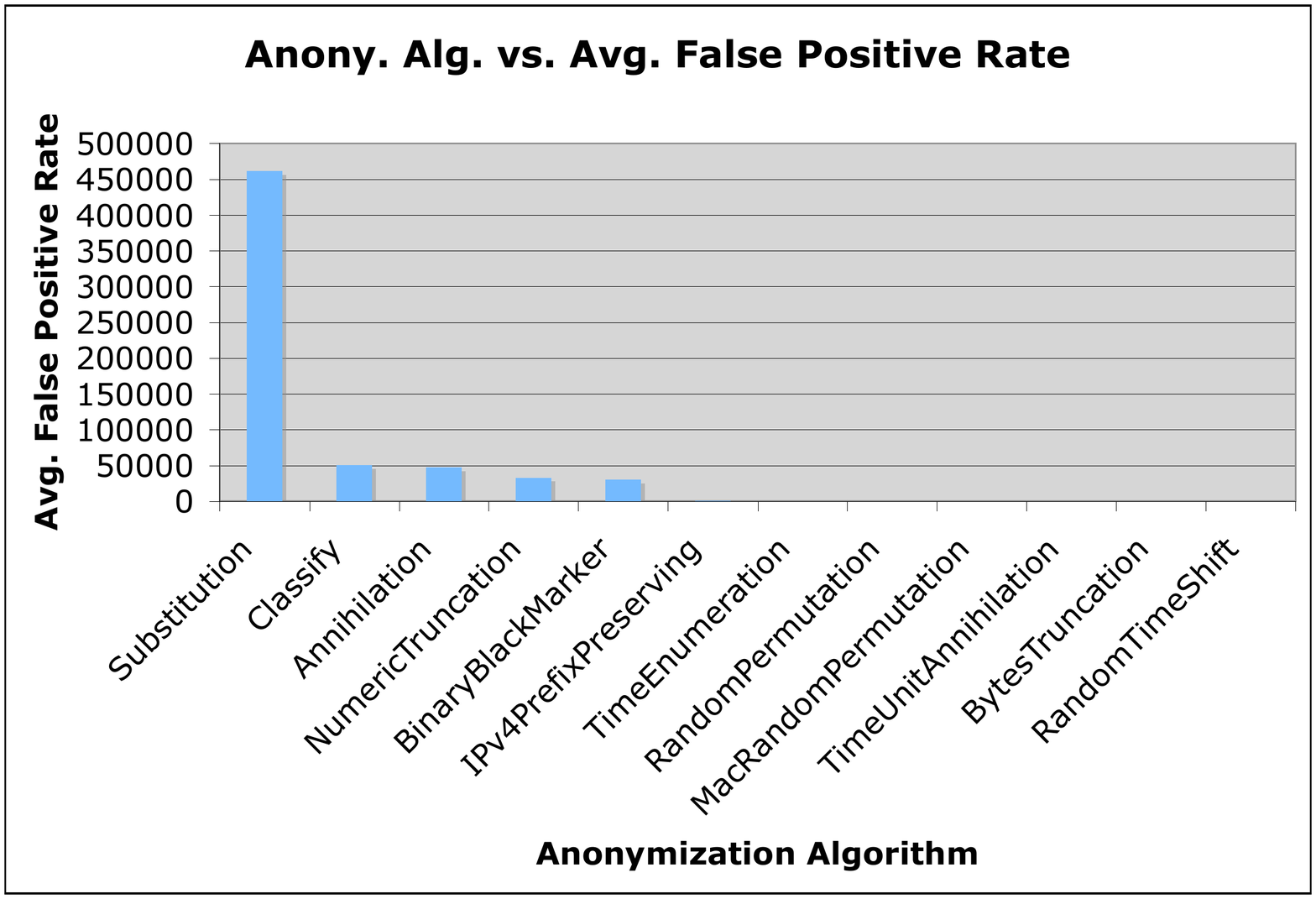} & \includegraphics[width=3in]{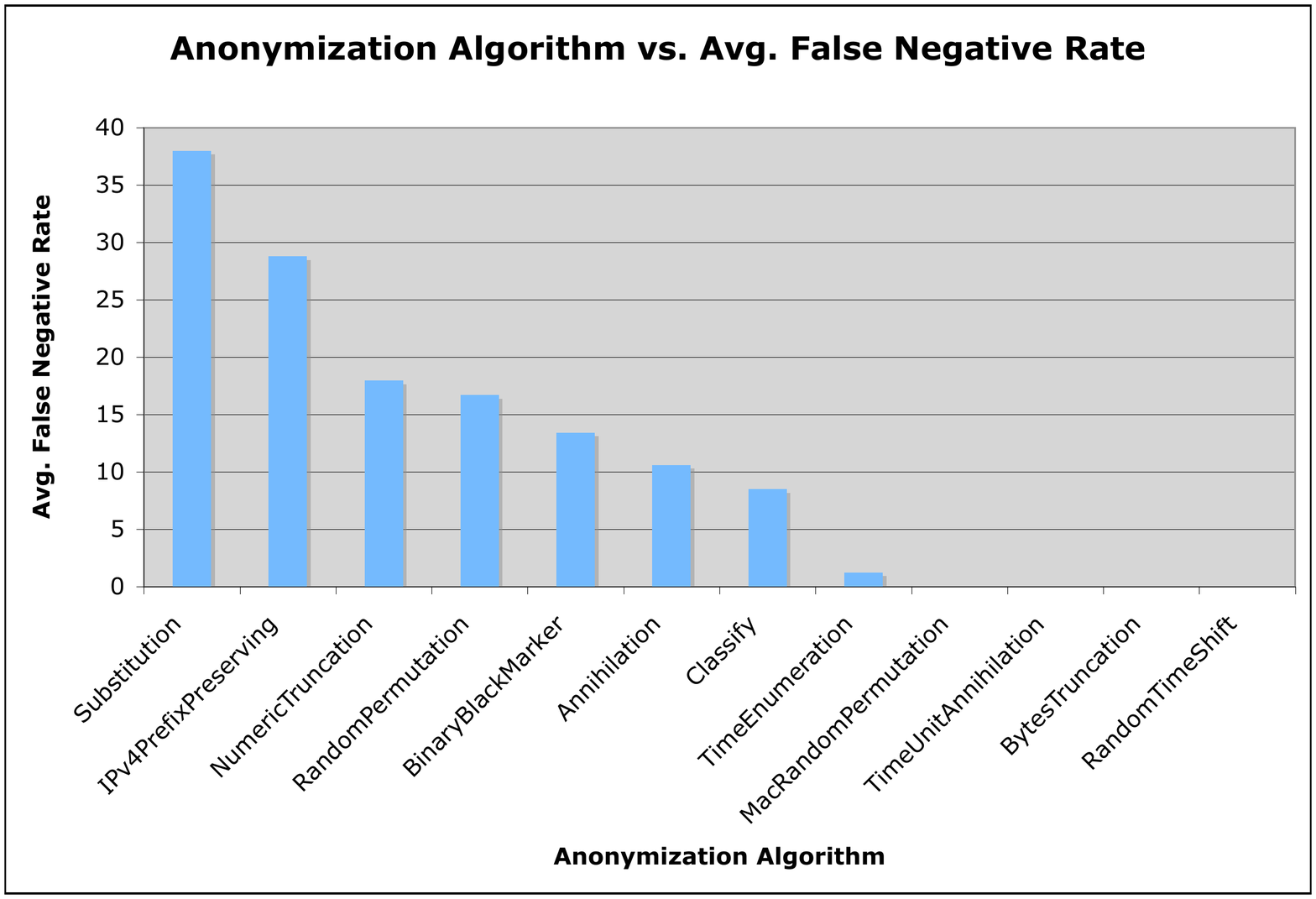} \\
  \end{tabular}
  \caption{The left hand chart shows the marginal of all the anonymization algorithms, with respect to false positives. The right hand chart shows the marginal of all anonymization algorithms with respect to false negatives.}
  \label{fig:marg-anonyvsfpfn}
\end{figure*}

Figure~\ref{fig:marg-fieldvsfpfn} shows the false positive and false negative marginals of the fields. The full data for these graphs is in Table~\ref{tab:marg-fieldFNFP}. Figure~\ref{fig:marg-anonyvsfpfn} shows the false positive and false negative marginals of the anonymization algorithms. The full data for these graphs is in Table~\ref{tab:marg-anonyFNFP}.

The fact that the majority of fields have no impact on utility, even
when anonymized in numerous ways, indicates that fields are more
important than the anonymization algorithm in determining the utility
of an anonymized log. If it was the other way around, we would expect
false positives and false negatives to be more evenly distributed over
all the fields.

Figure~\ref{fig:marg-fieldvsfpfn} indicates that the majority of
fields generate no false positives or false negatives. The fields for
which this is true (such as DST\_MAC) did not affect the utility of
the log under any anonymization algorithm. We can conclude that
anonymizing these fields has no impact on the utility of a log, with
respect to the IDS metric.

Consider the false positive rate first. We can see that very few
fields generated any false positives. This indicates that
anonymization, usually, does not add new patterns.

\subsection{Which fields have higher impact on utility?}
As Figure~\ref{fig:marg-fieldvsfpfn} and
Table~\ref{tab:marg-fieldFNFP} clearly show, the majority of fields do
not generate false positives. The fields that do are TCP\_DST\_PORT,
TCP\_SRC\_PORT, IPV4\_SRC\_IP, IPV4\_DST\_IP, TS\_SEC, and
TCP\_ACK\_NO.

We can see that several of the fields resulted in an average of 81
alerts. These fields (shown in italics in the table) had 0
error. Judging from these results, we can see that this set of fields
did not affect the utility of the log as measured by the IDS metric.

TCP\_DST\_PORT and TCP\_SRC\_PORT generated the most false alerts on
average. Upon inspection of the generated alert files, we can see that
most of the pairs generated only 2 types of alerts. In the case of
BinaryBlackMarker there were 520286 alerts of type 524. Alert 524 is ``BAD-TRAFFIC tcp port 0 traffic''. The other
anonymization algorithms produced the same pattern---the majority of
alerts were of type 524.

The reason for this is the value we substitute for the port field. The
BinaryBlackMarker, Substitution, NumericTruncation, and Annihilation
algorithms all replaced the port field with $0$. This resulted in the
alert being triggered for nearly all the packets in the log (See
Table~\ref{tab:anonyparam} for the parameter settings of the
anonymization algorithms). The same problem occurs for the
TCP\_SRC\_PORT field.

The RandomPermutation algorithm replaced the port number with another,
random port number, thus infrequently causing the generation of the BAD-TRAFFIC
alert. It is clear from this that the false positive rate was greatly
affected by the choice of the substitution port. However, the effect
would have been less pronounced had we counted the types of new alerts
rather than the raw alerts, themselves.

Arguably, it is the false negative count that is most important in
determining the utility of a log. A low false negative count indicates
that little information was lost in the process of anonymization. In terms
of false negatives, we find that there are $8$ fields that have an
impact on the average false negative rate:
TCP\_DST\_PORT,TCP\_FLAGS,TCP\_SRC\_PORT, IPV4\_DST\_IP, IPV4\_SRC\_IP,TCP\_ACK\_NO, TCP\_SEQUENCE, and TS\_SEC.

\subsection{What anonymization algorithms have higher impact on utility?}

Figure~\ref{fig:marg-anonyvsfpfn} summarizes the false positive and
false negative rates with respect to anonymization algorithm. For each
anonymization algorithm, the average false positive/false negative rate
is calculated over all the fields. Table~\ref{tab:marg-anonyFNFP}
contains the data for the graphs.

We can see from these that most anonymization algorithms have an
impact on the utility of a log. In contrast, the field data that we
saw before showed strong structure in what fields affected
utility. When viewed from the perspective of anonymization algorithms,
there is no single anonymization algorithm that stands out.

We can see from these that most anonymization algorithms have an
impact on the utility of a log. In contrast, the field data that we
saw before showed strong structure in what fields affected
utility. When viewed from the perspective of anonymization algorithms,
there is no single anonymization algorithm that stands out.

It might seem like the substitution algorithm has the largest effect
with a huge number of alerts. However, this is because of the
parameter setting used. By looking at the false negative rate we can
see that while substitution still has a large effect on utility, most
of the other algorithms have an effect as well.

\input{AnonyMarginal.tex} \input{FieldMarginal.tex}

%
%

    


    





%% file: rawAlertData.tex
\begin{table}[htp]
  \caption{Alerts generated for Anonymization-Field pairs. \textbf{AnonyAlg} is the anonymization algorithm used; \textbf{Field} is the field which it was applied on; \textbf{NumAlerts} is the number of alerts that were generated; \textbf{NumTypesOfAlerts} is the number of different \emph{types} of alerts generated. Table 1 of 2.}
  \label{table:raw-anony1}
  \begin{tabular}[t]{llll}
    \parbox{1in}{\textbf{AnonyAlg}} & \parbox{1in}{\textbf{Field}} & {\textbf{NumAlerts}} & \begin{minipage}{2in} \textbf{NumTypes} \\ \textbf{OfAlerts} \end{minipage} \\
    BinaryBlackMarker&SRC\_MAC&81&19 \\ 
    BytesTruncation&SRC\_MAC&81&19 \\ 
    Annihilation&SRC\_MAC&81&19 \\ 
    MacRandomPermutation&SRC\_MAC&81&19 \\ 
    BinaryBlackMarker&DST\_MAC&81&19 \\ 
    BytesTruncation&DST\_MAC&81&19 \\ 
    Annihilation&DST\_MAC&81&19 \\ 
    MacRandomPermutation&DST\_MAC&81&19 \\ 
    IPv4PrefixPreserving&IPV4\_SRC\_IP&1014&5 \\ 
    BinaryBlackMarker&IPV4\_SRC\_IP&9&4 \\ 
    Annihilation&IPV4\_SRC\_IP&9&4 \\ 
    RandomPermutation&IPV4\_SRC\_IP&9&4 \\ 
    NumericTruncation&IPV4\_SRC\_IP&9&4 \\ 
    IPv4PrefixPreserving&IPV4\_DST\_IP&759&5 \\ 
    BinaryBlackMarker&IPV4\_DST\_IP&9&4 \\ 
    Annihilation&IPV4\_DST\_IP&9&4 \\ 
    RandomPermutation&IPV4\_DST\_IP&14&5 \\ 
    NumericTruncation&IPV4\_DST\_IP&9&4 \\ 
    BinaryBlackMarker&IPV4\_ID&81&19 \\ 
    Annihilation&IPV4\_ID&81&19 \\ 
    NumericTruncation&IPV4\_ID&81&19 \\ 
    RandomPermutation&IPV4\_ID&81&19 \\ 
    Classify&IPV4\_ID&81&19 \\ 
    Annihilation&IPV4\_OFFSET&81&19 \\ 
    BinaryBlackMarker&IPV4\_TTL&81&19 \\ 
    Annihilation&IPV4\_TTL&81&19 \\ 
    NumericTruncation&IPV4\_TTL&81&19 \\ 
    RandomPermutation&IPV4\_TTL&81&19 \\ 
    Classify&IPV4\_TTL&81&19 \\ 
    Annihilation&IPV4\_CHECKSUM&81&19 \\ 
    BinaryBlackMarker&TCP\_DST\_PORT&520290&2 \\ 
    NumericTruncation&TCP\_DST\_PORT&457410&6 \\ 
    Substitution&TCP\_DST\_PORT&923033&2 \\ 
    Annihilation&TCP\_DST\_PORT&923033&2 \\ 
    RandomPermutation&TCP\_DST\_PORT&18&2 \\ 
    Classify&TCP\_DST\_PORT&521670&2 \\ 
    BinaryBlackMarker&TCP\_SRC\_PORT&380527&4 \\ 
    NumericTruncation&TCP\_SRC\_PORT&294519&6 \\
  \end{tabular}

\end{table}

\begin{table}[htp]
  \caption{Alerts generated for Anonymization-Field pairs. \textbf{AnonyAlg} is the anonymization algorithm used; \textbf{Field} is the field which it was applied on; \textbf{NumAlerts} is the number of alerts that were generated; \textbf{NumTypesOfAlerts} is the number of different \emph{types} of alerts generated. Table 2 of 2 }
  \label{table:raw-anony2}
  \begin{tabular}[t]{llll}
    \parbox{1in}{\textbf{AnonyAlg}} & \parbox{1in}{\textbf{Field}} & {\textbf{NumAlerts}} & \begin{minipage}{2in} \textbf{NumTypes} \\ \textbf{OfAlerts} \end{minipage} \\
    Substitution&TCP\_SRC\_PORT&922171&6 \\ 
    Annihilation&TCP\_SRC\_PORT&922171&6 \\ 
    RandomPermutation&TCP\_SRC\_PORT&20&4 \\ 
    Classify&TCP\_SRC\_PORT&381954&4 \\ 
    BinaryBlackMarker&TCP\_SEQUENCE&81&19 \\ 
    NumericTruncation&TCP\_SEQUENCE&65&15 \\ 
    Annihilation&TCP\_SEQUENCE&65&15 \\ 
    Classify&TCP\_SEQUENCE&81&19 \\ 
    BinaryBlackMarker&TCP\_ACK\_NO&62&16 \\ 
    NumericTruncation&TCP\_ACK\_NO&56&14 \\ 
    Annihilation&TCP\_ACK\_NO&56&14 \\ 
    Classify&TCP\_ACK\_NO&81&19 \\ 
    BinaryBlackMarker&TCP\_FLAGS&5&3 \\ 
    NumericTruncation&TCP\_FLAGS&5&3 \\ 
    Annihilation&TCP\_FLAGS&5&3 \\ 
    BinaryBlackMarker&TCP\_WINDOW&81&19 \\ 
    NumericTruncation&TCP\_WINDOW&81&19 \\ 
    Annihilation&TCP\_WINDOW&81&19 \\ 
    Classify&TCP\_WINDOW&81&19 \\ 
    Annihilation&TCP\_CHECKSUM&81&19 \\ 
    BinaryBlackMarker&TCP\_URGENT&81&19 \\ 
    NumericTruncation&TCP\_URGENT&81&19 \\ 
    Annihilation&TCP\_URGENT&81&19 \\ 
    Classify&TCP\_URGENT&81&19 \\ 
    Annihilation&TCP\_OPTIONS&81&19 \\ 
    BinaryBlackMarker&UDP\_DST\_PORT&81&19 \\ 
    NumericTruncation&UDP\_DST\_PORT&81&19 \\ 
    Substitution&UDP\_DST\_PORT&81&19 \\ 
    Annihilation&UDP\_DST\_PORT&81&19 \\ 
    RandomPermutation&UDP\_DST\_PORT&81&19 \\ 
    Classify&UDP\_DST\_PORT&81&19 \\ 
    BinaryBlackMarker&UDP\_SRC\_PORT&81&19 \\ 
    NumericTruncation&UDP\_SRC\_PORT&81&19 \\ 
    Substitution&UDP\_SRC\_PORT&81&19 \\ 
    Annihilation&UDP\_SRC\_PORT&81&19 \\ 
    RandomPermutation&UDP\_SRC\_PORT&81&19 \\ 
    Classify&UDP\_SRC\_PORT&81&19 \\ 
    Annihilation&UDP\_CHECKSUM&81&19 \\ 
    RandomTimeShift&TS\_SEC&81&19 \\ 
    TimeUnitAnnihilation&TS\_SEC&81&19 \\ 
    Annihilation&TS\_SEC&81&19 \\ 
    BinaryBlackMarker&TS\_SEC&83&20 \\ 
    TimeEnumeration&TS\_SEC&399&19 \\ 
    Annihilation&TS\_USEC&81&19 \\
  \end{tabular}
\end{table}

%% file: AnonyMarginal.tex
\begin{table}
  \caption{False negative/positive counts for an anonymization algorithm, aggregated over all log fields.}
  \label{tab:marg-anonyFNFP}
  \begin{tabular}{lll}
    \textbf{Anonymization Alg.} & \textbf{False Negatives} & \textbf{False Positive} \\
    Annihilation&10.62&47312.69 \\
    BinaryBlackMarker&13.43&30027.37 \\
    BytesTruncation&0&0 \\
    Classify&8.5&50200.83 \\
    IPv4PrefixPreserving&28.8&351 \\
    MacRandomPermutation&0&0 \\
    NumericTruncation&17.96&32692.13 \\
    RandomPermutation&16.72&2.11 \\
    RandomTimeShift&0&0 \\
    Substitution&38&461298.5 \\
    TimeEnumeration&1.25&80.75 \\
    TimeUnitAnnihilation&0&0 \\
  \end{tabular}
\end{table}

%% file: FieldMarginal.tex
\begin{table}
  \caption{False negative/positive counts for a field, aggregated over all anonymization algorithms}
  \label{tab:marg-fieldFNFP}
  \begin{tabular}{lll}
    \textbf{Field} & \textbf{False Negatives} & \textbf{False Positive} \\
    DST\_MAC&0&0 \\
    IPV4\_CHECKSUM&0&0 \\
    IPV4\_DST\_IP&72&151 \\
    IPV4\_ID&0&0 \\
    IPV4\_OFFSET&0&0 \\
    IPV4\_SRC\_IP&72&201 \\
    IPV4\_TTL&0&0 \\
    SRC\_MAC&0&0 \\
    TCP\_ACK\_NO&20&2.75 \\
    TCP\_CHECKSUM&0&0 \\
    TCP\_DST\_PORT&77.67&557572.33 \\
    TCP\_FLAGS&76&0 \\
    TCP\_OPTIONS&0&0 \\
    TCP\_SEQUENCE&8&0 \\
    TCP\_SRC\_PORT&75.5&483554.83 \\
    TCP\_URGENT&0&0 \\
    TCP\_WINDOW&0&0 \\
    TS\_SEC&1.2&65.2 \\
    TS\_USEC&0&0 \\
    UDP\_CHECKSUM&0&0 \\
    UDP\_DST\_PORT&0&0 \\
    UDP\_SRC\_PORT&0&0 \\
  \end{tabular}
\end{table}

%% file: multifieldresults.tex

\section{Multi-Field Policy Analysis}
\label{sec:multi-field-policy}
By multi-field anonymization policies, we are referring to
anonymization schemes that transform two or more fields in a log. The
bulk of this work has focused on single field policy
analysis. However, single field policy analysis will lay the
groundwork for understanding more complex multi-field policies.



\begin{table}
  \caption{This table shows that a multi-field policy has uncertain affects on the utility of a log}
  \label{tab:multifield}
  \begin{tabular}{ll}
    \textbf{Policy} & \textbf{Alerts} \\
    IPV4\_SRC\_IP-IPv4PrefixPreserving	& 1014 \\
    TCP\_SRC\_PORT-RandomPermutation & 7 \\
    TS\_SEC-TimeEnumeration & 399 \\
    IPV4\_SRC\_IP,TCP\_SRC\_PORT & 1016 \\
    IPV4\_SRC\_IP,TCP\_SRC\_PORT and TS\_SEC & 1010 \\
  \end{tabular}
\end{table}

Multi-field policies are difficult to analyze because the fields are
often related in subtle ways, not just very direct ways, such as
between the ACK and SEQ numbers. The anonymization of additional
fields certainly does not affect the total number of false positive
alerts in a linear way. For example, examine
Table~\ref{tab:multifield}.

Here, we see that anonymizing 3 fields separately produces 1014, 7 and
399 false alerts, respectively. If we anonymize the first 2 fields, we
get $1016\neq 1014+7$ alerts. This may not be too surprising. However,
if we do all 3 fields together, we get only $1010$ alerts---rather
than something around $1420=1014+7+399$. This is actually fewer false
alerts than any one field being anonymized in isolation. 

So while there is always more information loss when anonymizing more
fields, there may be a critical point at which one starts getting
fewer false positives as they do additional anonymization. We suspect
that this is always the case since complete anonymization will leave
nothing to alert upon. But clearly, more analysis must be done on
multi-field anonymization, and our future work will be focused on
evaluating multi-field anonymization policies.

%% file: related.tex
\section{Related Work}\label{related}
Work in data sanitization for computer and network logs to date has
focused almost entirely upon development of tools
\cite{slagell06flaim,li05canine,yurcik07anonyids,luo06scrubpa,foukarakis07flexible,Pang03}
and anonymization algorithms \cite{xu02prefixpreserving}, with little
to no work doing any formal analysis of the effects of
anonymization. This is an important aspect missing from the research
body since without it, we just have a lot of tools to haphazardly
anonymize data without knowing how to do it wisely or
effectively. This is also in stark contrast to k-anonymity
\cite{sweeney02kanonymity} which is usually applied to medical and
census type data and has received much more attention from the
research community.

While writing this paper, a new piece of work closely related to ours
appeared \cite{yurcik07anonyids}. Though this is mostly a paper
presenting yet another anonymization tool for pcap logs, at the end
they introduce a similar method of analyzing the effects of
anonymization of pcap traces by use of an IDS. Theirs is a cursory
analysis, only considering anonymization of one field at a time and
for only a few different fields. Also, their analysis only considered
the number of alerts, where the more alerts are generated the more
``security analysis'' is provided. Clearly, more than just the number
of alerts needs to be considered when evaluating utility, such as the
false positive and negative rates. We were unable to reproduce their
results (which may be because they used an unspecified subset of the
LBNL data set\footnote{http://www.icir.org/enterprise-tracing/}), but
more troubling is the use of this data set in the first place. To
evaluate the effects of anonymization, one must start with
unanonymized data. However, this LBNL data set is already
anonymized. So one has no clean baseline with which to compare the
anonymized data in this case.

%% file: conclusion.tex
\section{Conclusions}
\label{sec:conclusions}
Anonymization can be a powerful tool to allow greater cooperation between organizations. The need for cooperation is strikingly clear. However, a clear understanding of the needs of the data provider and the client is necessary before flexible, effective sharing between organizations can occur.

The objective of this paper has been to begin to formally evaluate the
utility vs. security trade off. The \emph{IDS metric} is simple yet
effective in evaluating the difference in utility when anonymizing
different fields in different ways.

In this paper, we have focused on answering three questions: whether
the field affects anonymization more than the algorithm; which fields
have a larger impact on utility; and which anonymization algorithms
have a larger impact on utility.

We have provided a thorough evaluation of single field anonymization
polices upon pcap formatted network traces. We found that the primary
impact on the utility of a log is not the particular anonymization
algorithm, but rather the field that was anonymized.

In addition, we were able to empirically show a range of utilities for
a log based on the field that was anonymized. The loss of utility was
largest for ports and IP addresses. There was some loss of utility for
the fields of ID, sequence number, flags, timestamp, and ACK
number. However, for many of the fields there was no change in utility
when anonymized.

This empirical evaluation provides the basis for further work on
studying the impact of more complex anonymization schemes on the
utility of a log.

\section{Future Work}
\label{sec:future-work}
There are numerous ways in which this work can be extended. First of
all, it is clear that evaluating utility via Snort generated IDS
alerts will cause the utility to depend upon the rule set. In fact, it
is as of yet unclear exactly how the rule set impacts the utility
measure. Though, we suspect there is a significant effect since we
found the specific fields anonymized has more effect than how it was
anonymized, and Snort rules usually focus on a one or two fields.

While we used Snort for all of our experiments, it is just one IDS and
only one type. It would be very interesting to investigate whether
anomaly-based IDSs are affected in similar ways.

The strength of an anonymization algorithm is a measure of how difficult it is to ``break'' or ``de-anonymize'' a log that has been anonymized via the algorithm. It is clear that we want strong anonymization algorithms so that attackers will have a difficult time to break the algorithm. As \cite{slagell06flaim} points out, there is a trade off between the security of an anonymization algorithm and the utility of the log. We have not discussed the strength of an anonymization algorithm in this work.

Our work is currently limited to anonymization policies for just one
field. Our next step would be to extend this work to multiple field
anonymization policies and to work with more realistic data. The
DARPA evaluation data set is useful because it is supervised, however
it is still synthetic. In the future, we will be gaining access to
other large {\em unanonymized} data sets that we can use instead of
the DARPA data. However, it is important for this research to have
unanonymized baseline sets.

Finally, we have still focused on utility for just one task, attack
detection. An incident responder does more than just detect attacks,
and in the future, we could look at how anonymizing logs affects other
important security related tasks---such as alert correlation.


%% file: appendix.tex
\section{Appendix}

\begin{table}[b!]
  \caption{Fields in a SNORT alert.}
  \label{tab:snort-alert-fields}
  \centering
  \begin{tabular}[t]{ccc}
    \begin{minipage}[t]{.33\linewidth}
      \begin{titemize}
      \item timestamp
      \item sig\_generator
      \item sig\_id
      \item sig\_rev
      \item msg
      \item proto
      \item src
      \item srcport
      \item dst
      \item dstport
    \end{titemize}
  \end{minipage} &
  \begin{minipage}[t]{.33\linewidth}
    \begin{titemize}
      \item ethsrc
      \item ethdst
      \item ethlen
      \item tcpflags
      \item tcpseq
      \item tcpack
      \item tcplen
      \item tcpwindow
      \item ttl
    \end{titemize}
  \end{minipage} &
  \begin{minipage}[t]{.33\linewidth}
    \begin{titemize}
      \item tos
      \item id
      \item dgmlen
      \item iplen
      \item icmptype
      \item icmpcoe
      \item icmpid
      \item icmpseq
    \end{titemize}
  \end{minipage}
\end{tabular}  
\end{table}

\begin{figure*}[t!]
  \includegraphics[width=7.0in]{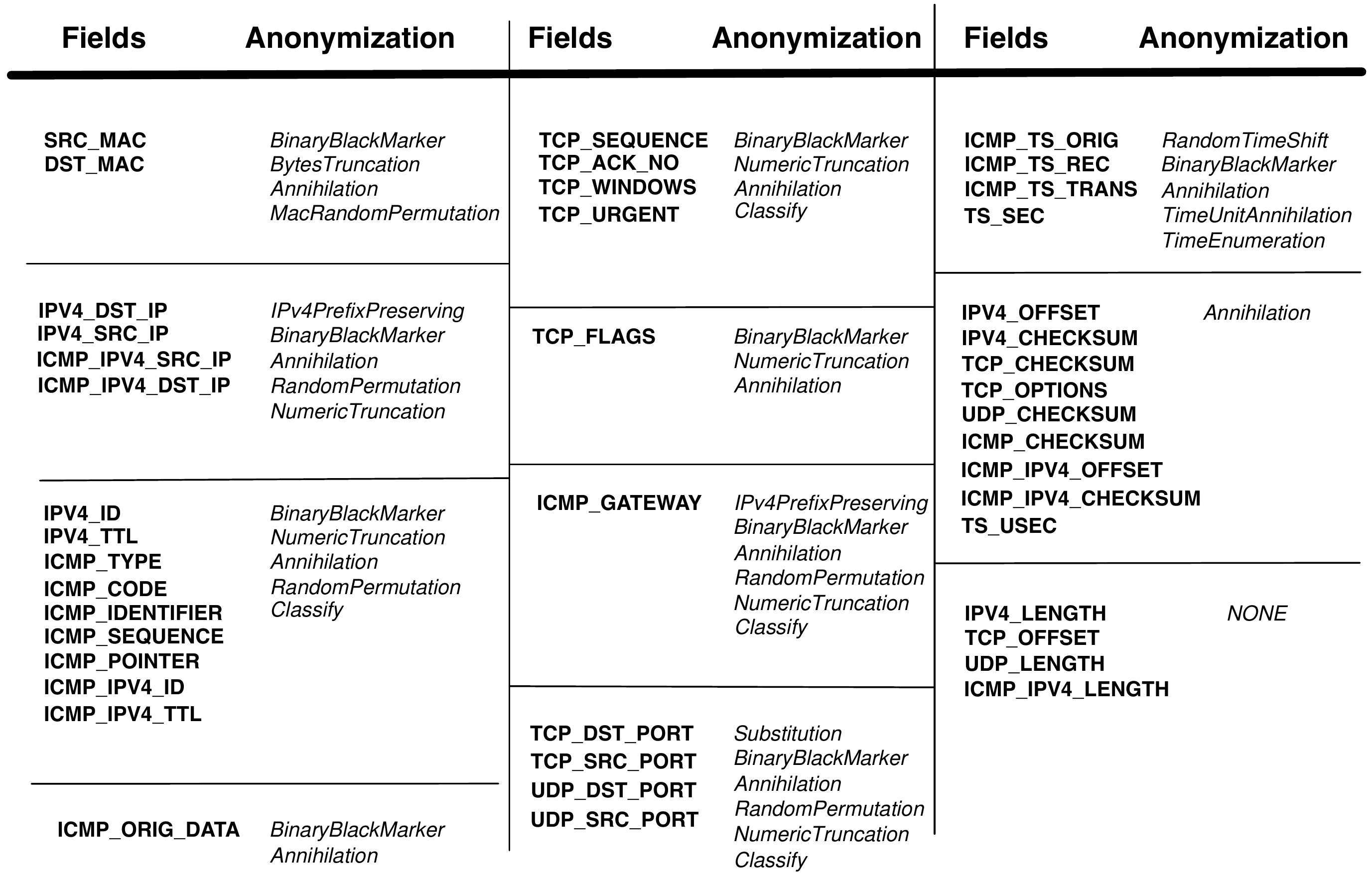}
  \caption{PCAP Fields and Anonymization Algorithms. Each section contains the fields (on the left) on which any of the anonymization algorithms on the right can be applied. For instance, Only the anonymization algorithms BinaryBlackMarker and Annihilation can be applied to the ICMP\_ORIG\_DATA field (the bottom left section)}
\label{fig:pcapanoy}
\end{figure*}

\input{./AnonyParameters.tex}

\begin{table*}[t!]
  \caption{Anonymization algorithms with applicable data types.}
  \label{tab:anony-algorithms}
  \begin{tabular}[t]{p{1.5in}|l|p{3in}}
    {\bf Anonymization Alg.} & {\bf Data Type(s)} & {\bf Description} \\ 
    Prefix-preserving & binary & Implements prefix-preserving permuta-\\
     & & tion described in \cite{xu02prefixpreserving}\\
    Truncation & binary & Removes suffix or prefix of data by\\
     & string & specified number of units.\\
     & numeric & \\
    Hash & binary & Outputs cryptographic has of data.\\
     & string & \\
     & hostname & \\
    Black Marker & binary & Overwrites specified number of units \\
     & string & with specified constant.\\
     & hostname & \\
    Time Unit Annihilation & timestamp & Annihilates particular time units\\
     & & (e.g., hour and minute units).\\
    Random Time Shift & timestamp & Randomly shifts timestamps within\\
     & & given window by same amount.\\
    Enumeration & timestamp & Preserves order, but not distance \\
     & & between elements.\\
    Random Permutation & binary & Creates random 1-to-1 mapping.\\
    Annihilation & binary & Replaces field with NULL value.\\
     & string & \\
    Classify & numeric & Partitions data into multiple non-\\
     & & overlapping subsets.\\
    Substitution & binary & Replaces all instances with a particu-\\
     & numeric & lar constant value.
  \end{tabular}
\end{table*}

%% file: AnonyParameters.tex
\begin{table*}[b!]
  \renewcommand{\arraystretch}{1.5}
  \centering
  \caption{Parameters to the Anonymization Algorithms}
  \label{tab:anonyparam}
  \begin{tabular}[c]{||llp{2in}||} 
    \hline
    Anonymization Algorithm & Parameters & Description \\ \hline
    & & \\
    IPV4PrefixPreserving &
    \begin{minipage}[t]{.4\linewidth}
      \begin{tdescription}
      \item[Passphrase:] foobar
      \end{tdescription}
    \end{minipage} &
    Sets the anonymization passphrase. \\ \hline

    MacRandomPermutation  &
    \begin{minipage}[t]{.4\linewidth}
      \begin{tdescription}
      \item[None]
      \end{tdescription}
    \end{minipage} &
    No Parameters. \\ \hline

    RandomTimeShift  &
    \begin{minipage}[t]{.4\linewidth}
      \begin{tdescription}
        \item[lowerTimeShiftLimit:] 16250000
        \item[upperTimeShiftLimit:] 31500000
      \end{tdescription}
    \end{minipage} &
    Shift time by a random amount between the lower and upper limits. \\ \hline

    TimeUnitAnnihilation  &
    \begin{minipage}[t]{.4\linewidth}
      \begin{tdescription}
      \item[timeField:] years
      \end{tdescription}
    \end{minipage} &
    Annihilate the years portion of the timestamp. \\ \hline

    NumericTruncation  &
    \begin{minipage}[t]{.4\linewidth}
      \begin{tdescription}
        \item[numShift:] 5
          \item[radix:] 2
      \end{tdescription}
    \end{minipage} &
    Shorten the field by 5 bits. \\ \hline

    TimeEnumeration &
    \begin{minipage}[t]{.4\linewidth}
      \begin{tdescription}
        \item[baseTime:] 0
        \item[intervalSize:] 1
      \end{tdescription}
    \end{minipage} &
    Set the time of the oldest record to 0. One timestp is equal to adding 1 to the timestamp field. \\ \hline

    RandomPermutation  &
    \begin{minipage}[t]{.4\linewidth}
      \begin{tdescription}
        \item[None]
      \end{tdescription}
    \end{minipage} &
    No Parameters. \\ \hline

    Annihilation  &
    \begin{minipage}[t]{.4\linewidth}
      \begin{tdescription}
        \item[None]
      \end{tdescription}
    \end{minipage} &
    No Parameters. \\ \hline

    Classify  &
    \begin{minipage}[t]{.4\linewidth}
      \begin{tdescription}
        \item[configString:] 1024:0,65536:65535
      \end{tdescription}
    \end{minipage} &
    Set elements less than 1024 to 0, and all others to 65535. \\ \hline
 
    BinaryBlackMarker  &
    \begin{minipage}[t]{.4\linewidth}
      \begin{tdescription}
        \item[numMarks:] 8
        \item[replacement:] 0
      \end{tdescription}
    \end{minipage} &
    Mark 8 bits as 0 \\ \hline

    BytesTruncation &
    \begin{minipage}[t]{.4\linewidth}
      \begin{tdescription}
        \item[numbits:] 20
        \item[direction:] left
      \end{tdescription}
    \end{minipage} &
    Remove 20 bits, starting from the left. \\ \hline

    Substitution &
    \begin{minipage}[t]{.4\linewidth}
      \begin{tdescription}
        \item[substitute:] 0
      \end{tdescription}
    \end{minipage} &
    Substitute field with 0. \\ \hline
  \end{tabular}

\end{table*}

%% file: arxiv08anonymization.bbl
\begin{thebibliography}{10}

\bibitem{honeynetChallenge}
Honeynet project challenge of the month.
\newblock http://www.honeynet.org/misc/chall.html, March 2003.

\bibitem{argus08website}
Argus: Qosient software.
\newblock http://www.qosient.com, March 2008.

\bibitem{bethencourt05mappinginternet}
J.~Bethencourt, J.~Franklin, and M.~Vernon.
\newblock Mapping internet sensors with probe response attacks.
\newblock In {\em Proceedings of the 14th USENIX Security Symposium}, August
  2005.

\bibitem{coull07playing}
S.~Coull, C.~V. Wright, F.~Monrose, M.~P. Collins, and M.~Reiter.
\newblock Playing devil's advocate: Inferring sensitive information from
  anonymized network traces.
\newblock In {\em Proceedings of the 14th Annual Network and Distributed System
  Security Symposium}, February 2007.

\bibitem{foukarakis07flexible}
M.~Foukarakis, D.~Antoniades, S.~Antonatos, and E.~P. Markatos.
\newblock Flexible and high-performance anonymization of netflow records using
  anontool.
\newblock In {\em Proceedings of the 3rd SECOVAL Workshop}, September 2007.

\bibitem{flaim-manual}
L.~W. Group.
\newblock Flaim core user's guide, March 2008.

\bibitem{aol06nytimes}
K.~Hafner.
\newblock Researchers yearn to use aol logs, but they hesitate.
\newblock New York Times, August 23 2006.

\bibitem{kohno05remotephysical}
T.~Kohno, A.~Broido, and K.~C. Claffy.
\newblock Remote physical device fingerprinting.
\newblock In {\em Proceedings of the IEEE Symposium on Security and Privacy},
  May 2005.

\bibitem{li05canine}
Y.~Li, A.~Slagell, K.~Luo, and W.~Yurcik.
\newblock Canine: A combined conversion and anonymization tool for processing
  netflows for security.
\newblock In {\em Proceedings of the 10th International Conference on
  Telecommunication Systems, Modeling and Analysis}, November 2005.

\bibitem{lincoln04privacypreserving}
P.~Lincoln, P.~Porras, and V.~Shmatikov.
\newblock Privacy-preserving sharing and correlation of security alerts.
\newblock In {\em Proceedings of the 13th USENIX Security Symposium}, August
  2004.

\bibitem{lippmann00darpa1999}
R.~Lippmann, J.~W. Haines, D.~J. Fried, J.~Korba, and K.~Das.
\newblock The 1999 darpa off-line intrusion detection evaluation.
\newblock {\em Computer Networks}, 2000.

\bibitem{lundin99privacy}
E.~Lundin and E.~Jonsson.
\newblock Privacy vs. intrusion detection analysis.
\newblock In {\em Proceedings of the 2nd International Workshop on Recent
  Advances in Intrusion Detection}, September 1999.

\bibitem{luo06scrubpa}
K.~Luo, Y.~Li, C.~Ermopoulos, W.~Yurcik, and A.~Slagell.
\newblock Scrub-pa: A multi-level, multi-dimensional anonymization tool for
  process accounting.
\newblock In {\em ACM Computing Research Repository (CoRR)}, January 2006.

\bibitem{mchugh00testing}
J.~McHugh.
\newblock Testing intrusion detection systems: A critique of the 1998 and 1999
  darpa intrusion detection system evaluations as performed by lincoln
  laboratories.
\newblock {\em ACM Transactions on Information and System Security},
  3(4):262--294, 2000.

\bibitem{narayanan06howtobreak}
A.~Narayanan and V.~Shmatikov.
\newblock How to break anonymity of the netflix prize dataset.
\newblock In {\em ACM Computing Research Repository}, October 2006.

\bibitem{pang06devil}
R.~Pang, M.~Allman, V.~Paxson, and J.~Lee.
\newblock The devil and packet trace anonymization.
\newblock {\em Computer Communication Review}, January 2006.

\bibitem{Pang03}
R.~Pang and V.~Paxson.
\newblock A high-level programming environment for packet trace anonymization
  and transformation.
\newblock In {\em Proceedings of the AMC SIGCOMM Conference}, August 2003.

\bibitem{seeberg07anewclassification}
V.~E. Seeberg and S.~Petrovic.
\newblock A new classification scheme for anonymization of real data used in
  ids benchmarking.
\newblock In {\em Proceedings of the 2nd International Conference on
  Availability, Reliability and Security}, April 2007.

\bibitem{sicker07legalissues}
D.~Sicker, P.~Ohm, and D.~Grunwald.
\newblock Legal issues surrounding monitoring during network research.
\newblock In {\em Proceedings of the 7th ACM SIGCOMM conference on Internet
  measurement}, August 2007.

\bibitem{slagell06flaim}
A.~Slagell, K.~Lakkaraju, and X.~Luo.
\newblock Flaim: A multi-level anonymization framework for computer and network
  logs.
\newblock In {\em Proceedings of the 20th USENIX Large Installation System
  Administration Conference}, December 2006.

\bibitem{slagell05sharingnetworklogs}
A.~Slagell, Y.~Li, and K.~Luo.
\newblock Sharing network logs for computer forensics: A new tool for the
  anonymization of netflow records.
\newblock In {\em Proceedings of the Computer Network Forensics Research
  Workshop}, August 2005.

\bibitem{slagell05networkloganonymization}
A.~Slagell, J.~Wang, and W.~Yurcik.
\newblock Network log anonymization: Application of crypto-pan to cisco
  netflows.
\newblock In {\em Proceedings of the NSF/AFRL Workshop on Secure Knowledge
  Management}, September 2004.

\bibitem{slagell05sharingcomputer}
A.~Slagell and W.~Yurcik.
\newblock Sharing computer network logs for security and privacy: A motivation
  for new methodologies of anonymization.
\newblock In {\em Proceedings of SECOVAL: The Workshop on the Value of Security
  through Collaboration}, August 2005.

\bibitem{sweeney02kanonymity}
L.~Sweeney.
\newblock K-anonymity: A model for protecting privacy.
\newblock {\em International Journal on Uncertainty, Fuzziness and
  Knowledge-based Systems}, 10, 2002.

\bibitem{vrable05scalability}
M.~Vrable, J.~Ma, J.~Chen, D.~Moore, E.~Vandekieft, A.~C. Snoeren, G.~M.
  Voelker, and S.~Savage.
\newblock Scalability, fidelity, and containment in the potemkin virtual
  honeyfarm.
\newblock In {\em Proceedings of the 20th Symposium on Operating System
  Principles}, October 2005.

\bibitem{xu02prefixpreserving}
J.~Xu, J.~Fan, M.~H. Ammar, and S.~B. Moon.
\newblock Prefix-preserving ip address anonymization: Measurement-based
  security evaluation and a new cryptography-based scheme.
\newblock In {\em Proceedings of the 10th IEEE International Conference on
  Network Protocols}, November 2002.

\bibitem{yurcik07anonyids}
W.~Yurcik and etc.
\newblock Scrub-tcpdump: A multi-level packet anonymizer demonstrating
  privacy/analysis tradeoffs.
\newblock In {\em Proceedings of the 3rd SECOVAL Workshop}, September 2007.

\bibitem{zalewskip0f}
M.~Zalewski.
\newblock p0f: Passive os fingerprinting tool.
\newblock http://lcamtuf.coredump.cx/p0f.shtml, March 2008.

\end{thebibliography}
